\documentstyle[psfig]{mn}

\def\ltsima{$\; \buildrel < \over \sim \;$}
\def\lsim{\lower.5ex\hbox{\ltsima}}
\def\gtsima{$\; \buildrel > \over \sim \;$}
\def\gsim{\lower.5ex\hbox{\gtsima}}
\def\mes{M\'esz\'aros}

\begin{document}

\title[Polarization of beamed gamma--gay bursts]
{Polarization lightcurves and position angle variation of beamed 
gamma--ray bursts}

\author[ Ghisellini \& Lazzati]
{Gabriele Ghisellini$^1$ \& Davide Lazzati$^{1,2}$\\
$^1$ Osservatorio Astronomico di Brera, Via Bianchi 46, I--23807
Merate (Lc), Italy \\
$^2$ Dipartimento di Fisica, Universit\`a degli Studi di Milano,
Via Celoria 16, I--20133 Milano, Italy \\
E--mail: {\tt gabriele@merate.mi.astro.it, lazzati@merate.mi.astro.it}}

\maketitle

\begin{abstract}
The recently detected linear polarization in the optical lightcurve of 
GRB~990510 renewed the interest on how polarization can be produced in 
gamma--ray burst fireballs.
Here we present a model based on the assumption that we are seeing a 
collimated fireball, observed slightly off--axis.
This introduces some degree of anisotropy, and makes it possible
to observe a linearly polarized flux even if the magnetic field
is completely tangled in the plane orthogonal to the line of sight.
We construct the lightcurve of the polarization flux, showing that
it is always characterized by two maxima, with the polarization position 
angle changing by $90^\circ$ between the first and the second maximum.
The very same geometry here assumed implies that the total flux
initially decays in time as a power law, but gradually steepens
as the bulk Lorentz factor of the fireball decreases.
\end{abstract}

\section{Introduction}
\label{uno}
It is now widely believed that the afterglow emission of
gamma--ray bursts is due to the deceleration of the relativistic
fireball in the circum--burst matter (for reviews see
Piran 1999; \mes~ 1999).
This produces a shock that accelerates electrons to random
relativistic energies and probably enhances the magnetic field, 
leading to the production of synchrotron emission.
If the magnetic field is completely tangled over the entire
emission seen by the observer, the resulting synchrotron emission 
is unpolarized.
On the other hand a very high degree of linear polarization can be
expected if a fraction of the magnetic field is well ordered,
reaching 60--70\% in the case of a completely ordered field.
Polarization values in the optical band in the range 3--30\% have been indeed 
observed in cosmic sources, like BL Lac objects and High Polarization
Quasars (see e.g. Angel \& Stockman 1980; Impey \& Tapia 1990),
whose radiation is believed to be produced by the synchrotron process.
One therefore expects that also in gamma--ray burst afterglows the
emission is polarized, and attempts were made to measure it.
After the upper limit (2.3\%) found for GRB~990123 (Hjorth et al. 1999),
Covino et al. (1999)  detected linear polarization in the afterglow 
of GRB~990510, at the small but significant level of 1.7$\pm$0.2\%.
This detection was then confirmed by Wijers et al. (1999)
who detected similar polarization values two hours and one day later.

On the theoretical side, Gruzinov \& Waxman (1999, hereafter GW99)
and Gruzinov (1999) predict values around 10\%, 
significantly larger than observed.
This estimate is based on the assumption that the overall
emission reaching the observer is produced in a finite number $N\sim 50$
of regions causally disconnected, each of which is embedded in a
completely ordered magnetic field. The predicted total polarization
level is $60\%/\sqrt{N}$, equal to $\sim$10\% for $N\sim 50$.
GW99 discuss how the coherence length of the magnetic field generated at 
the external shock front of a GRB fireball grows with time.
If, however, the magnetic field is generated at the collisionless
shock front, which is extremely local, it is not clear why the magnetic
field embedded in the newly swept matter should be linked to the field 
in the regions behind the shock.

An alternative magnetic field generation process (and hence geometry)
has been discussed by Medvedev \& Loeb (1999, ML99 hereafter), 
who consider a magnetic field
completely tangled in the plane of the shock front, but with a high degree
of coherence in the orthogonal direction. In the case of a spherical fireball
this geometry produces no polarization unless a part of the 
fireball emission is amplified and part is obscured, as is the case 
of interstellar scintillation.
In this case, however, the resulting polarization can be much better observed  
at radio wavelengths and should show a rapid and erratic change 
of the position angle. 

We here propose an alternative model, in which the magnetic field geometry
is analogous to that of ML99,\footnote{Note, however, that the ML99 instability
is not the only process that can be responsible of such a geometry, see e.g.
Laing 1980.} but in a fireball that is collimated in a cone and 
observed slightly off--axis. 
In this case the circular symmetry is broken and net polarization
can be observed (see e.g. Hjorth et al. 1999, Covino et al. 1999, 
Wijers et al. 1999). 
Evidences for beaming of the fireball of GRB~990510 from the anomalous 
decay of the optical lightcurve has been discussed in many recent papers 
(Harrison et al. 1999, Israel et al. 1999, Stanek et al. 1999b).

The key assumption of our model is that the fireball is collimated 
in a cone, observed slightly off--axis.
The key result we obtain is the polarization lightcurve, its connection 
with the flux behavior and a characteristic change of 90$^\circ$ in the 
polarization angle, making the model very testable.

\section{Polarization lightcurve}
\label{due}

\subsection{Magnetic field configuration}

Assume a slab of magnetized plasma, in which the configuration of the
magnetic field is completely tangled if the slab is observed face on,
while it has some some degree of alignment if the slab is observed edge on.
Such a field can be produced by compression in one direction of a volume
of 3D tangled magnetic field (Laing 1980, hereafter L80) or by Weibel 
instability (ML99).
If the slab is observed edge--on, the radiation is therefore polarized at a 
level, $P_0$, which depends on the degree of order of the field in the plane. 
At the angle $\theta$ from the normal of the slab, the degree of polarization 
can be expressed by, following L80:
\begin{equation}
{P(\theta)\over P_0} = {\sin^2(\theta) \over 1+\cos^2(\theta)}
\label{eq:polphi}
\end{equation}
If the emitting slab moves in the direction normal to its plane
with a bulk Lorentz factor $\Gamma$, we have to take into
account the relativistic aberration of photons.
This effect causes photons emitted at $\theta^\prime=\pi/2$ in the (primed) 
comoving frame $K^\prime$ to be observed at $\theta\sim 1/\Gamma$ 
(see also ML99).

\subsection{Polarization of beamed fireballs}

\begin{figure}
\psfig{figure=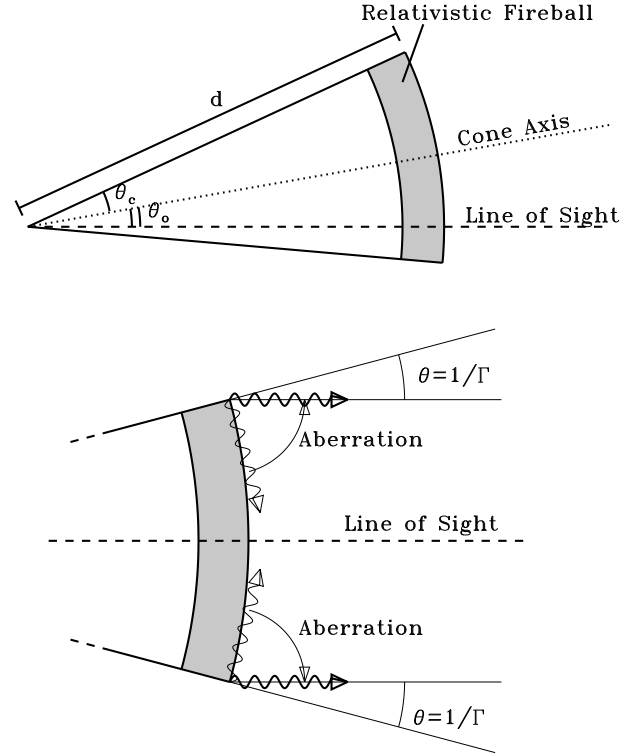,width=8.2cm}
\caption{{Geometry of the beamed fireball (upper panel). 
$\theta_o$ is the observer angle, $\theta_c$ is the cone aperture angle. 
The lower panel shows a zoom of the region around the line of sight. 
Note that the photons emitted in the comoving frame at an angle
$\pi/2$ from the velocity vector are those making an angle 
$\theta\sim 1/\Gamma$ with the line of sight in the observer frame.}
\label{fig:geom}}
\end{figure}

\begin{figure}
\psfig{figure=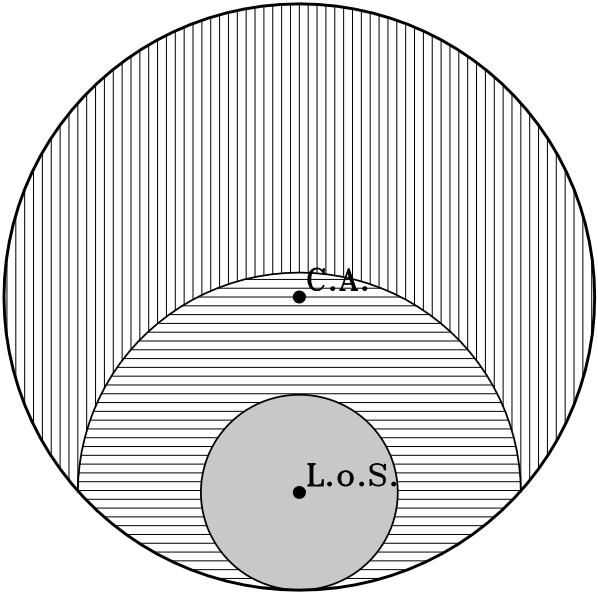,width=8.2cm}
\caption{{Front view of the beamed fireball, as in Fig. 1.
C.A. stands for Cone Axis, while L.o.S. stands for
Line of Sight.
The grey shaded area produces unpolarized radiation, due to 
the complete symmetry around the line of sight. 
The horizontal line shaded region 
produces a horizontal component of polarization while the upper region 
(vertically line shaded) produces vertical polarization. 
}\label{fig:front}}
\end{figure}

We assume that in gamma--ray burst fireballs the emitting region is a 
slab expanding radially and relativistically, 
compressed along the direction of motion.
We assume also that the fireball is collimated into a cone 
of semi--aperture angle $\theta_c$, and that the line of sight 
makes an angle $\theta_o$ with the jet axis (upper panel of 
Fig.~\ref{fig:geom}).
As long as $\Gamma>1/(\theta_c-\theta_o)$, the observer receives photons 
from a circle of semi-aperture angle $1/\Gamma$ around $\theta_o$
(i.e. within the grey shaded area of Fig.~\ref{fig:front}).
Consider the edge of this circle: radiation coming from each sector is highly
polarized, with the electric field oscillating in radial direction 
(see also ML99).
As long as we observe the entire circle, the configuration is
symmetrical, making the total polarization to vanish.
However, if the observer does not see part of the circle, some net 
polarization survives in the observed radiation. 
This happens if a beamed fireball is observed off--axis when
$1/(\theta_c+\theta_o)<\Gamma<1/(\theta_c-\theta_o)$.

The probability to observe a cone along its axis is vanishingly small, 
since it corresponds to a small solid angle; it is therefore more likely 
to observe the collimated fireball off--axis.
If the cone angle $\theta_c$ is small, the probability
$p(\theta_o/\theta_c)$ is approximately distributed as:
\begin{equation}
p\left({\theta_o \over \theta_c}\right) \propto 
{\theta_o \over \theta_c}; \quad 
\langle\theta_o\rangle = {2\over3} \theta_c
\end{equation}
where $\langle\theta_o\rangle$ is the average off--axis angle.

\begin{figure}[!t]
\psfig{figure=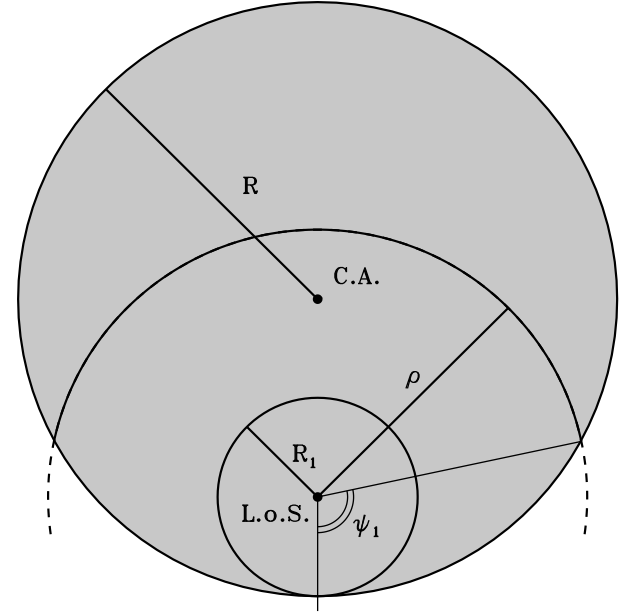,width=8.2cm}
\caption{{Sketch of the geometrical set up used to compute the
total and the polarized flux.
}
\label{fig:inte}}
\end{figure}

Assume therefore $\theta_o/\theta_c>0$ (Fig.~\ref{fig:front}). 
At the beginning of the afterglow, 
when $\Gamma$ is large, the observer sees only a small fraction of the fireball 
(grey shaded region in Fig.~\ref{fig:front}) and no polarization is observed.
At later times, when $\Gamma$ becomes smaller than $1/(\theta_c-\theta_o)$,
the observer will see only part of the circle centered in $\theta_o$:
there is then an asymmetry, and a corresponding net polarization flux
(horizontally line shaded region of Fig.~\ref{fig:front}).
To understand why the polarization angle in this configuration is horizontal, 
consider that the part of the circle which is not observed would have 
contributed to the polarization in the vertical direction. 
This missing fraction of vertical polarization does not cancel
out the corresponding horizontal one, which therefore survives.

At later times, as the fireball slows down even more, a larger area becomes 
visible. 
When $\Gamma\sim 1/(\theta_c+\theta_o)$, the dominant contribution
to the flux comes from the upper regions of the fireball
(see Fig.~\ref{fig:front}), which are vertically polarized.
The change of the position angle happens when the contributions from 
horizontal and vertical polarization are equal, resulting in a vanishing 
net polarization.
At still later times, when $\Gamma \to 1$, light aberration
vanishes, the observed magnetic field is completely tangled
and the polarization disappears.

We therefore expect two maxima in the polarization lightcurve,
the first for the horizontal component and the second for the 
vertical one.

The quantitative calculation of the lightcurve of the polarized fraction
has been worked out as follows. 
We assume that the emission of each small volume of the fireball is 
isotropic in the comoving frame $K^\prime$.
We also assume that in this frame each small element emits the same 
instantaneous intensity.
For each ring of radius $\rho$ and width $d\rho$ 
of fireball material around the 
line of sight (see Fig.~\ref{fig:inte}) the relativistically enhanced 
monochromatic intensity has been computed as 
$I(\nu) = \delta^3 I^\prime(\nu^\prime)$,
where $\delta\equiv [\Gamma(1-\beta\cos\theta)]^{-1}$ is the relativistic 
Doppler factor, and $\nu=\delta\nu^\prime$.
For $I^\prime(\nu^\prime)\propto (\nu^\prime)^{-\alpha}$ we then have
$I(\nu) = \delta^{3+\alpha} I^\prime(\nu)$.
Fig. 3 shows the geometrical set up of the system.
In our calculations,
the distances $R$, $R_1$ and $\rho$ are considered 
as angles, since all of them scale with the distance $d$ 
of the fireball from the center of explosion (see Fig. 1).
In this case, $\rho$ is equivalent to the angle $\theta$ between the
velocity vector of each element and the line of sight.
Therefore the intensity observed from each element 
is a function of $\Gamma$ and $\rho$.
The total intensity is obtained by integrating over the entire 
surface of the fireball, taking into account that each
ring is observed at a different angle, and then characterized by
a different $\delta$.
We assume a constant spectral index $\alpha=0.6$  
throughout our calculations.
The total intensity from the entire fireball at a given time
(and hence at a given $\Gamma$) is:
\begin{equation}
I(\Gamma,\nu) = 2\pi \int_0^{R_1} 
\!\!\! \!\!\!\! I(\Gamma,\nu,\rho) d\rho +
\int_{R_1}^{2R-R_1} 
\!\!\! \!\!\!\!\!\!\!\!\!\!\!\! 
I(\Gamma,\nu,\rho) \left[2\pi-2\psi_1(\rho)\right] d\rho \nonumber
\end{equation}
The first integral corresponds to the grey shaded area of Fig. 2,
and to $\rho<R_1$ (see Fig. 3).
The second integral corresponds to $\rho>R_1$.
In this case we receive radiation only from a sector of the ring,
between the angles $\psi_1$ and $2\pi-\psi_1$, with $\psi_1$ given by:
\begin{equation}
\psi_1\, =\, {\pi \over 2} - {\rm arcsin}\left[{ 2R_1R -R_1^2 -\rho^2 \over
2 \rho(R-R_1) }\right]
\end{equation}

We now compute the polarized intensity.
To this aim, we consider again each element of the ring of radius $\rho$. 
After calculating the corresponding viewing angle in the comoving frame,
we apply Eq. 1 to derive the intensity of the linearly polarized light.
The position angle of the polarization produced by 
each element is in the radial direction.
We then calculate the polarization of each ring by summing
the polarization vectors. Finally, we integrate over $\rho$.

It is convenient to 
write the polarization vector as a complex 
number\footnote{This is fully equivalent to the Stokes notation of the 
Q and U parameters for linear polarization.} 
$\vec{P}(\Gamma, \nu,\rho) \equiv 
P(\Gamma,\nu,\rho) e^{2i\theta_p}$ and integrate it between $\psi_1$ and 
$2\pi - \psi_1$ (see Fig.~\ref{fig:inte}). 
Here $\theta_p$ is the position angle of the linear polarization
in the observer frame.
The polarization of a generic ring is:
\begin{equation}
\vec{P}(\Gamma,\nu,\rho)\, =\, P(\Gamma,\nu,\rho)\,
\int_{\psi_1}^{2\pi-\psi_1}\!\!\!\!\!\!\!\!\!\!\!\!\!\!\! e^{2i\psi} \, d\psi 
=  P(\Gamma,\nu,\rho)\, \sin\left(2\psi_1\right) 
\end{equation}
Since the result is a real number, the polarization direction can lie 
either in the plane that contains 
both the line of sight and the cone axis (we call it ``vertical"
polarization angle) or in the orthogonal plane (i.e. ``horizontal"
polarization angle). No intermediate values are possible.

Integration over $\rho$ then yields:
\begin{eqnarray}
\vec{P}(\Gamma,\nu) \!\!
&=& \!\!\!{1 \over I(\Gamma,\nu)} \int_{R_1}^{2R-R_1}
\!\!\! \!\!\!\!\!\!\!\!\!\!\!\! \!\! I(\Gamma,\nu,\rho)
P(\Gamma,\nu,\rho) \, d\rho \, \int_{\psi_1}^{2\pi-\psi_1} 
\!\!\! \!\!\!\!\!\!\!\!\!\!\!\!  e^{2i\psi} \, d\psi \nonumber \\
&=& \!\!\!{1 \over I(\Gamma,\nu)} \int_{R_1}^{2R-R_1}
\!\!\! \!\!\!\!\!\!\!\!\!\!\!\!\!\! I(\Gamma,\nu,\rho)
P(\Gamma,\nu,\rho) \sin\left[2\psi_1(\rho)\right] \, d\rho
\label{eq:pola}
\end{eqnarray}

For simplicity, we neglected the light travel time effects introduced 
by the overall curvature of the emitting regions (see Fig.~\ref{fig:geom}),
approximating the emitting volume with a slab.
Since the degree of polarization of each element is divided by the total
intensity of the same element, we expect these effects to be small 
in our case.
Note that to include this effect requires to assign a 
specific relation between $\Gamma$ and the observed time $t$,
which can be different according to different models
(due to e.g. adiabatic vs radiative evolution and/or
gradients in the interstellar density).

\begin{figure}
\psfig{figure=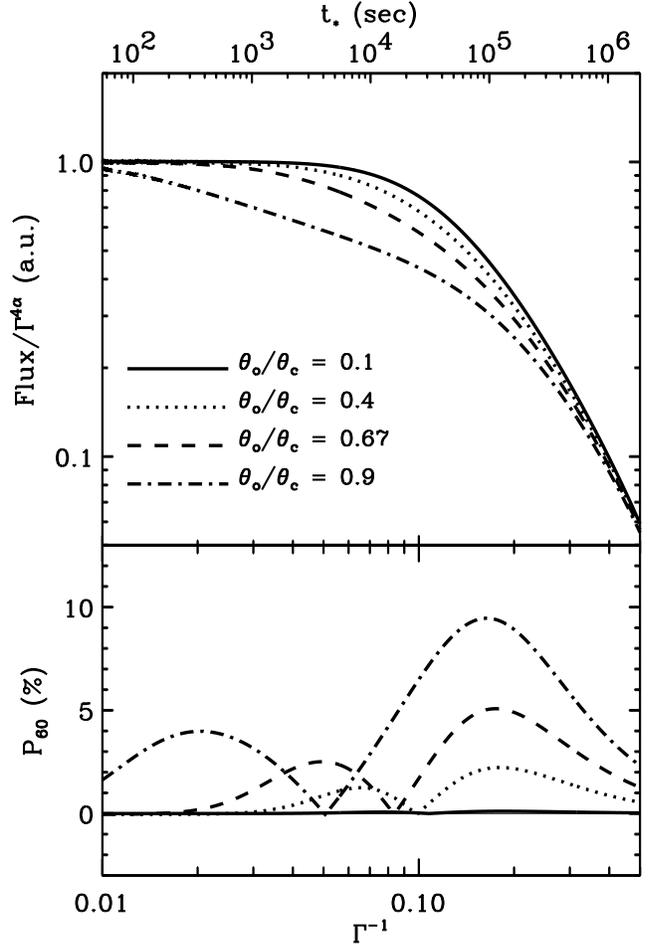,width=8.5cm}
\caption{{Lightcurves of the total flux (upper panel) and of the
polarized fraction (bottom panel) for four different choices
of the ratio $\theta_o/\theta_c$. 
The cone aperture angle $\theta_c=5^\circ$.
$\theta_o$ is the viewing angle, as defined in Fig.~\ref{fig:geom}.
The higher the ratio $\theta_o/\theta_c$, the higher the polarized fraction
due to the increase of the asymmetry of the geometrical setup.
The actual value of the observed polarization depends linearly
upon $P_0$ (see text). For this figure we assumed $P_0=60$\%.
The lightcurve of the total flux assumes a constant spectral index
$\alpha=0.6$ for the emitted radiation. 
Note that the the highest polarization values are associated with
total flux lightcurves steepening more gently.
To calculate the values of the upper x--axis (t$_\star$), we assumed
$(t_\star/t_0) = (\Gamma/\Gamma_0)^{-8/3}$ with
$t_0=50$ s and $\Gamma_0=100$.
}
\label{fig:lcur}}
\end{figure}

The result of the numerical integration of Eq.~\ref{eq:pola} is shown 
in the lower panel of Fig.~\ref{fig:lcur} for four  
different values of the off--axis ratio $\theta_o/\theta_c$. 
For the specific cases shown in the figure we have assumed 
$\theta_c = 5^\circ$, but the general properties
of the polarization lightcurve are unaffected by the particular choice 
of $\theta_c$.
All the lightcurves (except the one with the lower off--axis ratio,
which shows almost zero polarization), 
are characterized by two maxima in the polarized fraction. 
As discussed above, at the beginning the polarization is horizontal, 
it reaches a first maximum and then, when the polarized fraction 
vanishes, the angle abruptly changes by $\pi/2$. 
The polarized fraction raises again and then finally declines to zero at 
late times. 
The second peak has a value always larger than the first one.

By interpolating the results of the numerical integrations
(assuming $\alpha=0.6$), we can express the maximum of the 
polarization lightcurve as a function of the off--axis angle:
\begin{eqnarray}
P_{max} \simeq \!\!\!\!\!\!\! && 0.19 \,
P_0 \, \left( {\theta_o \over \theta_c} \right)^{2}; \label{eq:approx}\\
&& \left[{1\over 20} \le {\theta_o \over \theta_c} \le 1; \;\; 1^\circ \le 
\theta_c \le 15^\circ\right]  \nonumber
\end{eqnarray}
which is accurate within $5\%$ in the specified $\theta_o$ and $\theta_c$ 
ranges. 
This maximum always corresponds to the second peak.

\section{Link with the total flux lightcurve}
\label{tre}

The scenario proposed above for the polarization behavior has
a strict and direct link with the behavior of the lightcurve
of the total flux.
The exact connection will depend on the specific model
assumed for the deceleration of the fireball (i.e. homogeneous
or radially distributed density of the interstellar medium,
adiabatic or radiative regime, and so on), but there are
general properties that can be incorporated in any model.
For illustration, assume the simplest model of  
a spherical fireball expanding adiabatically in a homogeneous medium,
predicting that the bulk Lorentz factor $\Gamma\propto t^{-3/8}$,
and predicting a 
decay law for the flux density $F_\nu(t)\propto t^{-3\alpha/2}$
(\mes~ \& Rees 1997),
where $\alpha$ is the spectral index of the radiation spectrum, i.e.
$F_\nu\propto \nu^{-\alpha}$.
This law assumes that the flux is proportional
to the source solid angle: if the fireball has reached a region of size $d$,
then the accessible solid angle $\Omega \propto  (d/\Gamma)^2
\propto (t\Gamma)^2\propto t^{5/4}$.
In the case of a fireball collimated in a cone of constant opening angle
(e.g. neglecting for simplicity the possible lateral spreading
of the fireball: Rhoads 1997, 1999; see see also Moderski, Sikora 
\& Bulik 1999),
we will have the above relation as long as $1/\Gamma<\theta_c-\theta_o$.
On the other hand, when $1/\Gamma$ becomes greater than $\theta_c+\theta_o$,
all the cone front becomes visible, and therefore 
$\Omega \propto (\theta_c d)^2 \propto (\theta_c t \Gamma^2)^2 
\propto t^{1/2}$.
This produces a steepening of the power law lightcurve decay,
which now becomes $F_\nu(t) \propto t^{-3\alpha/2 - 3/4}$
(see also M\'esz\'aros \& Rees, 1999).
At intermediate times, for which 
$\theta_c-\theta_o< 1/\Gamma<\theta_c+\theta_o$,
the accessible solid angle increases with a law intermediate 
between $t^{5/4}$ and $t^{1/2}$, producing a $gradual$ steepening
in the lightcurve.

Assuming again $\Gamma\propto t^{-3/8}$, and calling
$\Gamma_1=1/(\theta_c-\theta_o)$, 
$\Gamma_2=1/(\theta_c+\theta_o)$ and $t_1$, $t_2$ the
corresponding times, we simply have
\begin{equation}
{\Gamma_1 \over \Gamma_2}\, =\, \left({t_2 \over t_1}\right)^{3/8}\, \to\,
{\theta_o\over \theta_c} \, =\, {(t_2/t_1)^{3/8}-1 \over
(t_2/t_1)^{3/8}+1}
\end{equation}
In the case of the GRB~990510 afterglow, the 
decay laws have been $t^{-0.9}$ at early times (Galama et al. 1999)
up to $\sim$ half a day after the burst event, 
and $t^{-2.5}$ after $\sim$5 days (Israel et al. 1999).
Therefore the ratio $t_2/t_1$ is of the order of 10--15,
implying $\theta_o/\theta_c$ between 0.4 and 0.45.
However, the observed steepening was larger than the value (3/4)
derived above.
Another cause of steepening can be the lateral spreading of
the fireball, as suggested by Rhoads (1997), when $\Gamma\sim 1/\theta_c$
(but see Panaitescu \& \mes~1999, who suggest that this phase should
occur later).
In addition, some steepening of the lightcurve decay 
may be due to a curved synchrotron spectrum:
in fact at early times the spectral index derived on the basis of $BVRI$
photometric observations, de--reddened with $E_{B-V}=0.20$, 
was flat ($\alpha=0.61\pm 0.12$, 21.5 hours after the burst,
Stanek et al. 1999b).
Such a flat spectral index in the optical band must necessarily
steepen at higher frequencies, to limit the emitted power.
An estimate of such a steepening will come from the analysis
of the X--ray afterglow flux, observed from 8 to 44.5 hours after 
the burst (Kuulkers et al. 1999).

We conclude that the observed steepening of the lightcurve decay 
may then be the combined result of a curved synchrotron spectrum 
and of a collimated fireball.
In this scenario, the spectral index of the late (after 5 days)
optical spectrum should be $\alpha\sim$ 1.1--1.2.

\section{Discussion}
\label{qua}
We have proposed a model to derive the amount of
linear polarization observable from a collimated fireball.
We have shown that some degree of polarization can be observed even if the
magnetic field is completely tangled in the plane of the fireball,
as long as the fireball is observed slightly off--axis.
One of the main virtues of the proposed model is in its
easily testable predictions:
i) the lightcurve of the degree of polarization has two maxima;
ii) the observable polarization position angles are fixed between
the first and the second maximum, being orthogonal between
each other; 
iii) there is a strong link with the lightcurve of the total flux.

The degree of polarization is predicted to be moderate,
reaching 10\% only if we are observing a collimated fireball
at its edge, and only for a short period of time.
A larger degree of polarization would then suggest that the
magnetic field is not completely tangled in the plane orthogonal to
the line of sight, as suggested by Gruzinov \& Waxman (1999).
Up to now very few attempts have been done to measure linear polarization
in optical afterglows, and is therefore premature to draw
any firm conclusion from the upper limit detected in GRB~990123
(Hjorth et al. 1999) and from the positive detection in the case
of GRB~990510 (Covino et al. 1999, Wijers et al. 1999).
Note however that the lightcurve behavior of GRB~990510
matches our predictions, as well as the constant position angle
of the observed polarization.
At the time of the two first polarization measurements of GRB 990510
(made two hours apart) the lightcurve of the total flux was decaying
as $t^{-1.3}$ (Stanek et al. 1999a), i.e. it was already steepening,
in agreement with our model.
Unfortunately, the polarization value measured one day later (Wijers et al.
1999) was not precise enough to further constrain the proposed scenario.

It will be very interesting to explore in the future the association of 
a gradually steepening lightcurve and the presence of polarization.
We cannot exclude the possibility that some of the already
observed afterglows can indeed be fitted by steepening power laws, 
but that the lack of data makes such an attempt meaningless.
In addition, there are some optical afterglows (e.g. GRB 980326,
Groot et al., 1998; GRB 980519, Halpern et al., 1999)
which showed a rapid decay.
In these cases we may have observed only that part of the lightcurve
corresponding to $1/\Gamma>\theta_c+\theta_o$. From 
these considerations we conclude that GRB~990510 may not be unique
in its category and that a large fraction of gamma--ray burst
afterglows can have some degree of optical linear polarization.

\end{document}